\documentclass{aastex}
\usepackage{spr-astr-addons}
\usepackage{url}

\usepackage[dvipsnames]{xcolor}
\usepackage{newtxtext,newtxmath}
\usepackage[T1]{fontenc}
\usepackage{ae,aecompl}
\usepackage{adjustbox}

\usepackage{graphicx}	
\usepackage{amsmath}	
\usepackage{amssymb}	

\begin{document}

\title{AU~Pegasi revisited: period evolution and orbital elements of 
a peculiar Type~II Cepheid}
\shorttitle{Period evolution of AU~Pegasi}
\shortauthors{G. Cs\"ornyei and L. Szabados}

\author{G\'eza Cs\"ornyei \altaffilmark{1}} \and 
\author{L\'aszl\'o Szabados \altaffilmark{2}}
\affil{Konkoly Observatory, Research Centre for Astronomy and Earth Sciences, Hungarian Academy of Sciences, Konkoly Thege Mikl\'os \'ut 15-17,\\ 1121 Budapest, Hungary\\
e-mail: csornyei.geza@csfk.mta.hu}
\email{csornyei.geza@csfk.mta.hu} 

\altaffiltext{1}{Institute of Physics, E\"otv\"os Lor\'and University, P\'azm\'any P\'eter s\'et\'any 1/a, Budapest 1117, Hungary}
\altaffiltext{2}{MTA CSFK Lend\"ulet Near-Field Cosmology Research Group}

(Submitted: 16th July; Accepted: 2nd September)

\begin{abstract}
New analysis on the period changes of Type~II Cepheid AU~Peg is presented. 
The available recent photometric measurements were collected and analysed 
with various methods. The period has been found to be constant for certain 
time intervals, although increasing in overall, in contrast with the 
previous expectations, which suggested the period change to reverse. 
Superimposed on overall period change, a formerly unknown periodic behaviour 
has been found in the $O-C$ diagram of AU~Peg, which cannot be matched to 
the radial velocity variations.
Since the Cepheid is a member of a binary system, it is probable that 
the unusual period change is in connection with the companion's tidal force. 
The orbital elements of the binary system involving AU~Peg have been also
revised.
\end{abstract}


\section{Introduction} \label{sec:intro}

AU~Pegasi is a Type II Cepheid with a pulsation period of approximately 
2.4 days and with a mean spectral type of F8. This Cepheid is considered 
unique in several ways; most importantly it has been found to have a 
physical companion (\cite{Harris1979}) and a highly unstable pulsation 
period (\cite{Szabados1977}, \cite{Erleksova1978}). The orbital period of 
this spectroscopic binary system is 53.3 days, which is the second shortest
among the known binaries with a Type~II Cepheid component. The only Type~II 
Cepheid in a binary system with a shorter orbital period is TYC 1031 01262 1
($P_{\textrm{orb}}$=51.38 days) (\cite{Antipin2007}), while the other 
Galactic Type II Cepheids in binary systems, IX~Cas and TX~Del have an 
orbital period of 110.29 and 130.15 days, respectively (Harris \& Welch 1989).

\cite{Harris1984} found that the colour of AU~Pegasi is unusually red, 
which would be normal for a Cepheid with a longer pulsation period. 
The effective temperature of its atmosphere is remarkably low, 
T$_{\textrm{eff}}=5500-6000$ K, depending on the pulsational phase 
(\cite{Kovtyukh2018}). Recent spectroscopic measurements revealed that 
the [Fe/H] abundance ratio is 0.27, which means that AU~Peg is a metal 
rich Type~II Cepheid (\cite{Kovtyukh2018}).

The infrared excess and the unusual colour index ($B-V$=0.85, 
\cite{Harris1984}, \cite{Wallerstein2002}) also indicate the presence of a 
dust cloud surrounding the binary system (\cite{McAlary1986}). It has also 
been observed that the spectrum of this Cepheid exhibits P~Cygni like 
behaviour, narrow emission features on the red side of the H$\alpha$ line, 
which show variations on orbital period timescale, thus might be a result 
of interaction between the atmosphere of the star with the circumstellar 
matter around it (\cite{Vinko1998}). Presently it is thought that AU~Pegasi 
is close to filling its Roche lobe and mass transfer between the two 
stars almost certainly occurred formerly (\cite{Maas2007}).

The temporal behaviour of the pulsation period of AU~Pegasi was 
extensively studied by \cite{Vinko1993} in the time interval of
JD\,2\,433\,100$-$2\,448\,600. They found that the pulsation period was 
slightly increasing with a rate of $dP/dt = 5\cdot 10^{-7} $ day/day 
before JD\,2\,440\,000, while the average pulsation period length was 
$P_{\textrm{pul}} = 2.39$ days. According to their study the first epoch 
where the rate of the period variation changed was between JD\,2\,439\,000 
and 2\,441\,000. At this epoch the rate of the variation accelerated 
to $dP/dt = 1.8\cdot 10^{-6}$ day/day. This period variation has 
eventually seemed to stop between JD\,2\,446\,700 and 2\,447\,800. 
After this second break point the period variation seemed to reverse and 
to start decreasing.

They concluded that the rapid period change might be the result of the
interaction between the variable star and its companion, but as they 
pointed out, the period variation cannot be explained by tidal interaction
alone. Since the classification of Type~II Cepheid is somewhat uncertain, 
it also had been suggested, that this variable star is a classical 
Cepheid crossing the instability strip for the first time (\cite{Vinko1993}).
 This could explain the rapid period variation, but the latest studies still 
 classified this variable as a Type~II Cepheid (\cite{Groenewegen2018}), 
which classification is also supported by the kinematics, the position of 
the star on the colour-magnitude diagram and length of the orbital period, 
as well.

The latest Gaia parallax of the object is $\Pi = 1.6739 
\pm 0.0448$ milliarcsecond (\cite{Gaia1}, \cite{Gaia2}), while the $V$-band 
extinction of the star is $A_V = 0.184$ mag (NASA/IPAC Infrared Science Archive), 
which together correspond to an absolute magnitude of $M_V = 0.069$ mag. 
According to the classical Cepheid period-luminosity relation (\cite{Benedict}) 
this would correspond to a period of 0.214 days, which is significantly smaller 
than the one we observed, thus it supports the classification of the star as a 
Type~II Cepheid.

In view of its importance and peculiarities, we extended the former studies 
with more recent photometric data covering the last 25 years. Our aim was 
to gain a better insight into the effect of orbital revolution on the 
pulsation period. 

\section{Observational data}

In order to determine the temporal variation of the pulsation period, 
photometric data that have been acquired after the last extensive study 
(\cite{Vinko1993}) were collected and analysed.
 The complete dataset 
contains measurements from All Sky Automated Survey (ASAS, 
\cite{Pojmanski2002}), All Sky Automated Survey for Supernovae 
(ASAS-SN,\\ \cite{Shappee2014}), Kamogata Sky Survey (KWS,\\ \cite{Morokuma2014}),  \emph{Hipparcos} (\cite{Perryman1997}) and \emph{Gaia} (\cite{Gaia1},
\cite{Gaia2}).

\begin{table}[!h]
\small
\begin{center}
\begin{tabular}{l c  c}
\hline
Source & HJD interval & $N$ \\
\hline
ASAS & 2\,452\,754 - 2\,455\,157 & 364\\
ASAS-SN & 2\,456\,389 - 2\,458\,380 & 1074\\
\emph{Hipparcos} & 2\,447\,889 - 2\,448\,973 & 75\\
KWS & 2\,455\,752 - 2\,458\,360 & 509\\
\emph{Gaia} & 2\,457\,164 - 2\,457\,390 & 14\\
\hline
\end{tabular}
\caption{Temporal information of the various surveys used for the analysis.}
\label{tab:mes.data}
\end{center}
\end{table}

\begin{table}[!t]
\begin{adjustbox}{width=\columnwidth,center}
\begin{tabular}{c c c c c c c}
\hline
JD$-$2\,400\,000 & $V$ & $B-V$ && JD$-$2\,400\,000 & $V$ & $B-V$ \\
\hline
49538.451 &    9.260 &    1.060 &&  51757.390 &    9.350 &    0.960 \\ 
49569.531 &    9.240 &    0.910 &&  51758.373 &    9.080 &    0.840 \\ 
49570.415 &    9.370 &    0.930 &&  51759.368 &    9.170 &    1.150 \\ 
49606.326 &    9.390 &    0.910 &&  51782.343 &    9.080 &    0.820 \\ 
49606.397 &    9.380 &    0.900 &&  51838.266 &    9.120 &    0.900 \\ 
49630.240 &    9.390 &    0.950 &&  51839.245 &    9.400 &    1.080 \\ 
49630.332 &    9.370 &    0.980 &&  51840.265 &    9.070 &    0.820 \\ 
49631.235 &    9.080 &    0.750 &&  51878.264 &    9.280 &    0.900 \\ 
49631.299 &    9.070 &    0.760 &&  52147.372 &    9.270 &    0.960 \\ 
49666.208 &    9.270 &    0.920 &&  52150.512 &    9.350 &    0.940 \\ 
49666.259 &    9.290 &    0.990 &&  52151.335 &    9.070 &    0.850 \\ 
49688.211 &    9.360 &    1.010 &&  52194.372 &    9.110 &    0.810 \\ 
49690.200 &    9.210 &    0.950 &&  52195.299 &    9.170 &    0.970 \\ 
49900.448 &    9.330 &    1.000 &&  52196.292 &    9.360 &    0.930 \\ 
49918.413 &    9.080 &    0.790 &&  52197.283 &    9.090 &    0.850 \\ 
49919.431 &    9.330 &    0.920 &&  52198.295 &    9.350 &    0.980 \\ 
49920.425 &    9.230 &    0.760 &&  52199.336 &    9.050 &    0.810 \\ 
49921.428 &    9.150 &    0.960 &&  52200.265 &    9.200 &    0.950 \\ 
49952.412 &    9.120 &    0.820 &&  52589.239 &    9.380 &    0.960 \\ 
49986.433 &    9.060 &    0.930 &&  52618.265 &    9.390 &    0.920 \\ 
49987.313 &    9.350 &    0.990 &&  52619.199 &    9.070 &    0.850 \\ 
49992.273 &    9.430 &    0.930 &&  52620.209 &    9.220 &    0.920 \\ 
50015.263 &    9.070 &    0.880 &&  52901.384 &    9.070 &    0.860 \\ 
50016.285 &    9.400 &    0.970 &&  52902.330 &    9.300 &    0.980 \\ 
50018.263 &    9.270 &    0.980 &&  52903.349 &    9.130 &    0.850 \\ 
50338.429 &    9.100 &    0.950 &&  52904.313 &    9.160 &    0.950 \\ 
50371.274 &    9.220 &    0.830 &&  52905.395 &    9.350 &    0.950 \\ 
50376.366 &    9.090 &    0.780 &&  52906.316 &    9.080 &    0.860 \\ 
50605.476 &    9.100 &    0.820 &&  52947.226 &    9.050 &    0.840 \\ 
50633.482 &    9.320 &    1.000 &&  52947.303 &    9.060 &    0.860 \\ 
50634.425 &    9.130 &    0.770 &&  52948.225 &    9.300 &    0.980 \\ 
50749.268 &    9.330 &    0.990 &&  52948.304 &    9.350 &    1.000 \\ 
50749.333 &    9.400 &    0.920 &&  53266.337 &    9.290 &    0.980 \\ 
50750.274 &    9.110 &    0.790 &&  53267.328 &    9.130 &    0.850 \\ 
50750.333 &    9.110 &    0.780 &&  53286.317 &    9.290 &    0.910 \\ 
50751.246 &    9.150 &    0.960 &&  53331.235 &    9.270 &    0.990 \\ 
50751.308 &    9.180 &    0.910 &&  53569.430 &    9.100 &    0.890 \\ 
51051.396 &    9.190 &    0.820 &&  53612.350 &    9.070 &    0.800 \\ 
51052.363 &    9.140 &    0.920 &&  53614.347 &    9.280 &    0.890 \\ 
51080.300 &    9.200 &    0.810 &&  54389.298 &    9.110 &    0.830 \\ 
51756.463 &    9.180 &    0.920 &&  54390.270 &    9.270 &    0.970 \\ 
\hline
\end{tabular}
\end{adjustbox}
\caption{$BV$ photometric data obtained with the 50 cm Cassegrain telescope 
at Piszk\'estet\H{o} Mountain Station of the Konkoly Observatory.}
\label{tab:Piszkes.data}
\end{table}

Table~\ref{tab:mes.data} contains information about the 
temporal coverage of the photometric surveys and number of observations 
analysed in this paper.
In addition, $BV$ photometric measurements obtained with the 50~cm 
Cassegrain telescope of the Piszk\'estet\H{o} Mountain Station of the 
Konkoly Observatory by one of us (L.~Sz.) between 1994 and 2007 were 
also involved in our study. Between 1994 and 1998, an integrating 
photoelectric photometer equipped with an unrefrigerated EMI 9058QB  
photomultiplier tube was attached to the telescope, while from the year 
2000 on an electrically cooled (to $-20^\circ$C) photon counting photometer 
containing an EMI 9203QB photomultiplier was mounted in the Cassegrain focus.

Both photometers were equipped with standard filters of the Johnson 
photometric system. The brightness of AU~Peg was observed using 
BD~+17$^\circ$~4575 as the comparison star (SIMBAD magnitudes: 
$V = 9.24$ mag; $B-V = 1.13$ mag), and BD~+18$^\circ$~4788 served as the 
check star. Table \ref{tab:Piszkes.data} contains these previously 
unpublished measurements obtained in the Piszk\'estet\H{o} Mountain Station.

Radial velocity (RV) measurement data have also been collected from the 
literature. In addition to the measurements made by \cite{Harris1984}, 
three additional sets of RV data have been available: those obtained by 
\cite{Barnes1988}, \cite{Gorynya1998} and \cite{Vinko1998}.

\section{Analysis of the photometric data}

\begin{figure}[!h]
\centering
\hspace*{-13pt}
\includegraphics[scale=1]{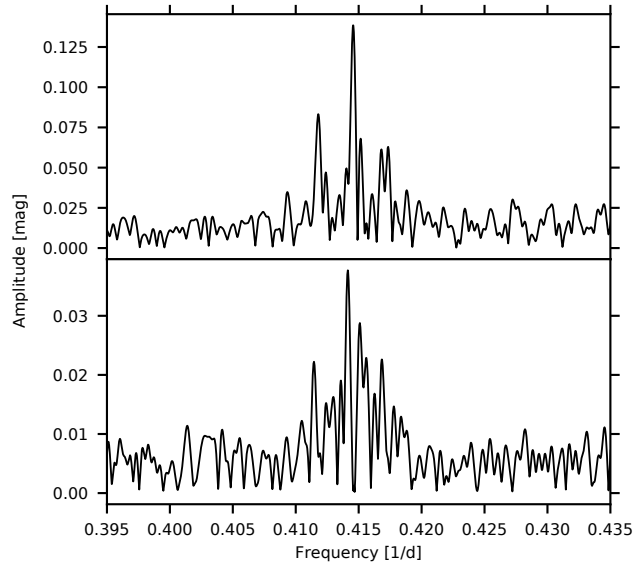}
\caption{The Fourier spectrum of the original ASAS data (top panel) 
and that of the residual data, after subtracting the main frequency 
(bottom panel).}
\label{fig:asas-four}
\end{figure}

Three different approaches have been used to obtain the period length 
of the pulsation for different time intervals: the discrete Fourier 
transformation (DFT, \cite{Deeming1975}), for which we have used the 
Period04 analysis software (\cite{Lenz2005}), the phase dispersion 
minimization (PDM, \cite{Stellingwerf1978}) and the method of $O-C$ diagrams 
(\cite{Sterken2005}). Since the first two methods require a constant 
or slowly varying pulsation period and for the construction of the $O-C$ 
diagram we would need the correct determination of the phase, e.g. the 
moments of a chosen phase ($O$) and the elapsed number of cycles ($E$), 
which would become uncertain in the case of strong period variation, 
the collected data had to be divided into shorter time intervals. 
The demonstration of this problem can be seen in Fig.~\ref{fig:asas-four}, 
where the Fourier amplitude diagrams of the entire ASAS data set of AU~Peg 
is presented. The highest peak on the upper panel corresponds to the 
pulsation period of $P_{\textrm{pul}}$ = 2.4122 days. The lower panel is 
obtained from the residuals after the fitting of the first frequency. 
The highest peak in this diagram corresponds to $P = 2.4147$ days. 
From the proximity of these period values we assumed that the Cepheid 
underwent a rapid period variation in the time interval covered by the 
ASAS observations.

\begin{figure}[!h]
\centering
\hspace*{-13pt}
\includegraphics[scale=1]{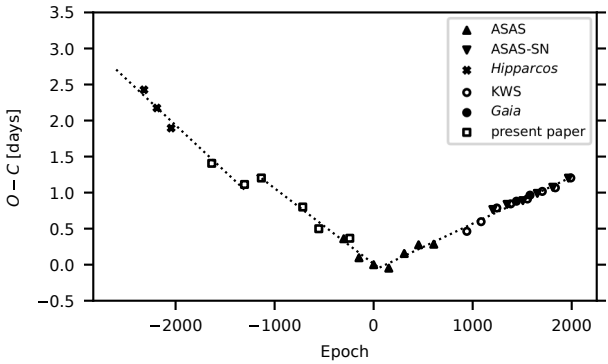}
\caption{The $O-C$ diagram of AU~Pegasi calculated from the new measurements 
with linear fit segments.}
\label{fig:au_o-c}
\end{figure}

Since the period variation covered by the collected 
data was not strong enough for the phase shift affecting the moments of
light curve extrema to accumulate into a longer time than the period itself, 
we decided to create the $O-C$ diagram for these measurements. 
As a first step of the analysis, every set of observation 
has been split into smaller subsets. For each survey, 250 day long temporal 
bins were defined, in which each datapoint was moved to a new subset. 
 The phase curves of each previously created subset of measurements have been calculated, which then were used to determine the observed ($O$) epoch values. This method inevitably decreases the resolution of the 
resulting $O-C$ curve, but the precision of the results increases, since the 
error of the phase calculation will decrease significantly. The $O-C$ diagram 
of $V$ band measurements was calculated assuming the elements:
$$ C = \textrm{JD}\,2\,453\,481.812 + 2.412 ^\textrm{d}\cdot E. $$
The obtained $O-C$ values are listed in Table \ref{tab:au_tab_oc}, while 
the corresponding diagram is presented in Fig.~\ref{fig:au_o-c}.

\begin{figure}[!h]
\centering
\hspace*{-13pt}
\includegraphics[scale=1]{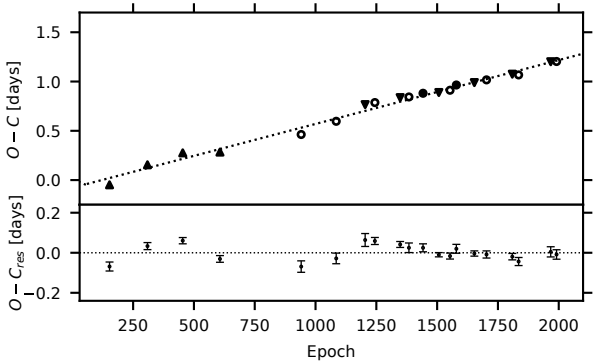}
\caption{Top panel: The last linear segment of the $O-C$ diagram and its 
linear fit. Bottom panel: The residual values of the $O-C$ diagram after 
subtracting the linear fit. The notation of the data points is the same 
as in Fig.~\ref{fig:au_o-c}.}
\label{fig:au_o-c_dev}
\end{figure}

It has been found that the segments of the $O-C$ graph can be described with 
linear functions, thus the pulsation period of the Cepheid remains approximately 
the same for the different time intervals. The change of the pulsation period 
shown by the ASAS data can be approximated as a parabolic function on the 
$O-C$ diagram, hence it can be described as a linear period change (see later 
in this chapter). Table \ref{tab:fit_params} describes the linear fits in the 
different time intervals. As illustrated in Fig.~\ref{fig:au_o-c_dev}, it has 
been found that the $O-C$ data points obtained from ASAS, ASAS-SN and KWS 
measurements deviate from the fitted linear curve in a periodic manner.

\begin{table}[!h]
\small
\begin{center}
\begin{tabular}{c c c}
\hline
Epoch & Linear fit & Period [d] \\
\hline
$-2600<E<-1280$ & $-0.0013\cdot E -0.0636$ & 2.4109 \\ 

 & \hspace*{1pt}$\pm 0.00013$ \hspace*{9pt} $\pm 0.0851$ &  \hspace*{-4pt}$\pm 0.0001$\\
 
\hspace*{-14pt}$-1180<E<60$ & $-0.0010\cdot E +0.0108$ & 2.4111 \\

 & \hspace*{1pt}$\pm 0.00014$ \hspace*{9pt} $\pm 0.0562$ & \hspace*{-4pt}$\pm 0.0001$ \\
 
\hspace*{6pt}$60<E<2000$ & \hspace*{5pt}$0.0006 \cdot E -0.0780$ & 2.4128\\

 & \hspace*{1pt}$\pm 0.00002$ \hspace*{9pt} $\pm 0.0243$ & \hspace*{-4pt}$\pm 0.0001$ \\
\hline
\end{tabular}
\caption{The fitted lines for the different time intervals (see 
Fig.~\ref{fig:au_o-c}), and the calculated periods.}
\label{tab:fit_params}
\end{center}
\end{table}

\begin{figure}[!h]
\centering
\hspace*{-13pt}
\includegraphics[scale=1]{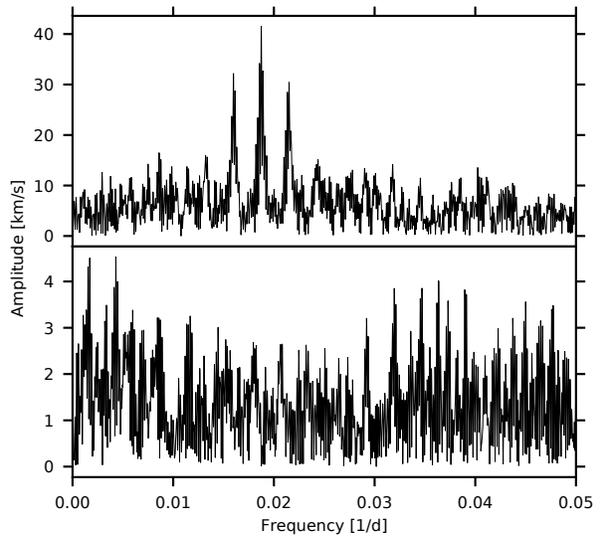}
\caption{The Fourier spectrum obtained from the RV measurements (top panel) 
and that of the residual data after whitening with the frequency of the 
known orbital motion.}
\label{fig:radvel}
\end{figure}

The period of this cyclic behaviour is approximately 2215 days, while its 
amplitude is 0.05 days. This variation cannot originate from the light-time 
effect caused by the known companion of the Cepheid, since the expected 
amplitude of this variation would be as low as 0.001 days, and the period 
is not appropriate, either. To examine whether the obtained periodic variation 
could correspond to the light-time effect of a formerly unknown companion, 
we analysed the available RV observations collected from the literature. 
The Fourier spectra of the RV observations before and after subtracting the 
main frequency (the known orbital motion) are presented in 
Fig.~\ref{fig:radvel}. Since the expected RV projection corresponding to 
the obtained period and amplitude of the variation is 45.7 km/s (assuming circular orbit), which would then result in a sharp 
peak in the Fourier diagram of the RV observations at the frequency of 
$4.515\cdot 10^{-4}$ cycles/day, that is clearly not present (although the 
Fourier spectrum shows a peak with a much smaller amplitude at that frequency), 
we cannot attribute the observed variation to any effect caused by the orbital 
motion of the Cepheid.

\begin{table}[!h]
\centering
\scalebox{0.88}{
\begin{tabular}{c c c c c}
\hline
HJD$-$2\,400\,000 & E & $O-C$ & $\sigma$ & Survey \\
& & (d) & (d) & \\ 
\hline
47888.061 & $-$2319 & 2.427 & 0.016 & \emph{Hipparcos} \\   
48201.388 & $-$2189 & 2.174 & 0.020 & \emph{Hipparcos} \\   
48548.458 & $-$2045 & 1.895 & 0.019 & \emph{Hipparcos} \\   
49536.949 & $-$1635 & 1.407 & 0.031 & present paper \\   
49898.720 & $-$1485 & 1.356 & 0.032 & present paper \\   
51755.516 & $-$716  & 0.800 & 0.060 & present paper \\   
52145.983 & $-$554  & 0.499 & 0.039 & present paper \\   
52753.705 & $-$302  & 0.360 & 0.021 & ASAS \\   
52900.852 & $-$241  & 0.367 & 0.043 & present paper \\   
53124.911 & $-$148  & 0.096 & 0.026 & ASAS \\   
53481.812 & 0       & 0.000 & 0.014 & ASAS \\   
53850.823 & 153     & \hspace*{-5pt}$-$0.048 & 0.022 & ASAS \\   
54227.321 & 309     & 0.155 & 0.018 & ASAS \\   
54577.203 & 454     & 0.276 & 0.015 & ASAS \\   
54946.269 & 607     & 0.284 & 0.016 & ASAS \\   
55752.104 & 941     & 0.462 & 0.028 & KWS \\   
56099.587 & 1085    & 0.597 & 0.026 & KWS \\   
56386.801 & 1204    & 0.766 & 0.032 & ASAS-SN \\   
56483.308 & 1244    & 0.787 & 0.018 & KWS \\   
56734.221 & 1348    & 0.837 & 0.015 & ASAS-SN \\   
56821.065 & 1384    & 0.843 & 0.024 & KWS \\   
56961.008 & 1442    & 0.881 & 0.019 & \emph{Gaia} \\
57117.804 & 1507    & 0.888 & 0.011 & ASAS-SN \\   
57228.786 & 1553    & 0.912 & 0.015 & KWS \\   
57291.555 & 1579    & 0.965 & 0.020 & \emph{Gaia} \\   
57470.078 & 1653    & 0.989 & 0.013 & ASAS-SN \\   
57590.713 & 1703    & 1.067 & 0.018 & KWS \\   
57846.458 & 1809    & 1.074 & 0.025 & ASAS-SN \\   
57909.167 & 1835    & 1.067 & 0.020 & KWS \\   
58227.704 & 1967    & 1.201 & 0.024 & ASAS-SN \\   
58285.598 & 1991    & 1.203 & 0.024 & KWS \\   
\hline
\end{tabular}
}
\caption{$O-C$ values of AU~Peg calculated from the seasonal $V$ band 
datasets of the different surveys.}
\label{tab:au_tab_oc}
\end{table}

\begin{figure}[!h]
\centering
\hspace*{-13pt}
\includegraphics[scale=1]{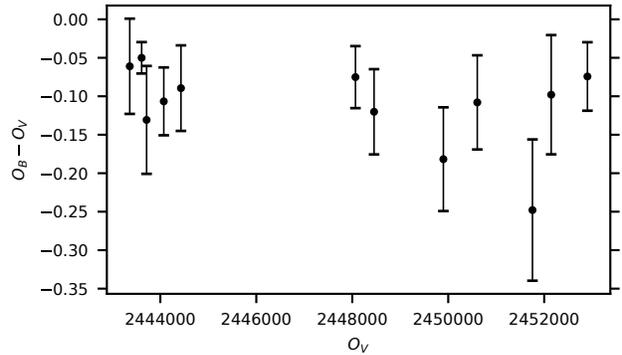}
\caption{The differences of the $O$ values in $B$ and $V$ bands in days 
as a function of the Julian Day.}
\label{fig:BV}
\end{figure}

In the case of most archival photometric data series and the measurements 
presented in this paper as well, not only $V$ band, but $B$ band 
observations were also available. With the use of available $B$ band data, 
another set of $O$ values has been calculated. Since these measurements 
covered the time interval in which a rapid period increase was observed 
(\cite{Vinko1993}), the $O-C$ diagram of $B$ band could not have been 
created, but the simultaneous $V$ and $B$ observations allowed the comparison 
of $B$ and $V$ band epochs. The differences of the same phase $O$ 
values calculated from $B$ and $V$ band observations are presented in 
Fig.~\ref{fig:BV}. Table \ref{tab:BV_data} contains the calculated 
differences of the two sets of $O$ values. According to the results, 
the brightness maximum in the $B$ band light curve precedes the $V$ band 
with approximately 0.082 days ($\Delta \phi$ = 0.039 for the phase shift). 
This corresponds well to the former observations (Freedman 1988), where a 
systematic phase shift was found for several Cepheids, which appeared 
to be increasing for longer wavelengths.
All photometric measurements have also been analysed with the DFT and 
PDM methods. To prevent the mixing of different period values, observation 
subsets shorter than 250 days were created. Data from different surveys 
were treated separately. For the PDM method parameters $N_{\textrm{b}}$=10 
and $N_{\textrm{c}}$=3 have been used, where $N_{\textrm{b}}$ and 
$N_{\textrm{c}}$ denote the number of bins and the number of covers 
(\cite{Stellingwerf1978}). 

\begin{figure}[!h]
\centering
\hspace*{-13pt}
\includegraphics[scale=1]{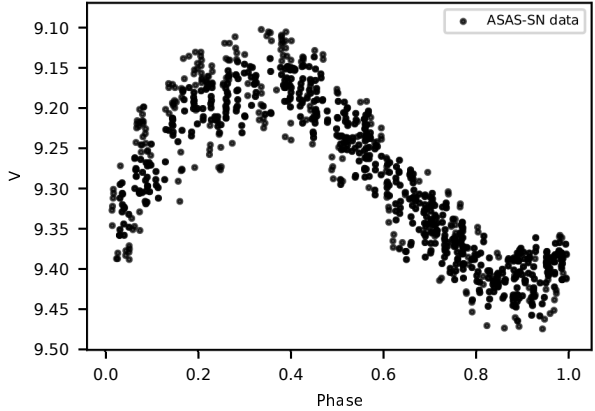}
\caption{Light curve in V calculated from the ASAS-SN data. The phases 
were calculated using the periods listed in Table~\ref{tab:period_tab}.}
\label{fig:phase_c}
\end{figure}

Some of the datasets, like the ASAS-SN observations, 
proved to contain singificant number of outliers and exhibit a higher 
scatter in the data, which would result in less precise period evaluation. 
To address this problem during the PDM fit, the outlying data points 
deviating from the calculated phase curve with more than $2\sigma$ 
were excluded, and the phase curve was calculated again. We have tested the 
method with different thresholds and the $2\sigma$ cut appeared to be the 
best choice for an automated outlier removal: with the $1\sigma$ threshold 
many points were removed that could not have been flagged as definite outliers, 
while at $3\sigma$ not all outliers were removed. An example for the $V$ band 
phase curve of the Cepheid is presented in Fig.~\ref{fig:phase_c}.

\begin{table}[!h]
\begin{center}
\scalebox{0.88}{
\begin{tabular}{c c c c}
\hline
HJD$-$2\,400\,000 & $O_{\textrm{B}}-O_{\textrm{V}}$ & $\sigma$ & Reference \\
\hline
43362.532 &     $-$0.061 &      0.062 & Vink\'o et~al. (1993) \\ 
43610.115 &     $-$0.050 &      0.021 & Henden (1980)\\ 
43713.625 &     $-$0.131 &      0.070 & Vink\'o et~al. (1993) \\ 
44071.742 &     $-$0.107 &      0.044 & Moffett (1984)\\ 
44430.211 &     $-$0.089 &      0.056 & Moffett (1984)\\ 
48066.403 &     $-$0.075 &      0.041 & Vink\'o et~al. (1993)\\ 
48454.503 &     $-$0.120 &      0.055 & Vink\'o et~al. (1993)\\ 
49898.720 &     $-$0.182 &      0.067 & present paper\\ 
50605.330 &     $-$0.108 &      0.061 & present paper\\ 
51755.516 &     $-$0.248 &      0.092 & present paper\\ 
52145.983 &     $-$0.098 &      0.078 & present paper\\ 
52900.852 &     $-$0.074 &      0.045 & present paper\\ 
\hline
\end{tabular}
}
\caption{ The time difference of the moments of brightness maxima and 
its standard deviation ($\sigma$) in different bands in days.}
\label{tab:BV_data}
\end{center}
\end{table}

\begin{figure}[!h]
\centering
\hspace*{-13pt}
\includegraphics[scale=1]{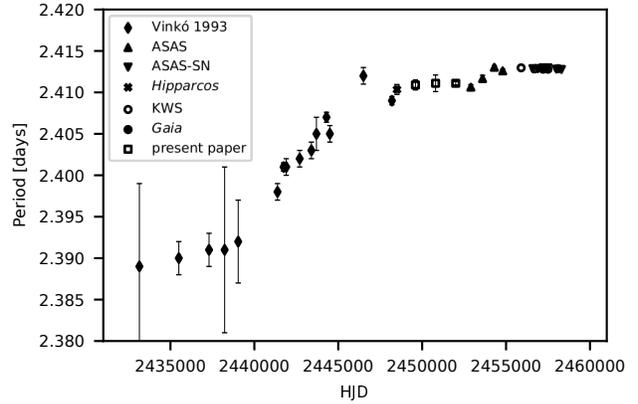}
\caption{The pulsation period of AU~Pegasi as a function of time}
\label{fig:periods}
\end{figure}

Table~\ref{tab:period_tab} contains the calculated period values for the 
different surveys and various methods. The period has been calculated 
with every method, if the temporal coverage and the amount of data points 
in the seasonal subset allowed it. For the ASAS, ASAS-SN and KWS the period 
has been calculated with every method. In the case of the \emph{Gaia} data, 
the amount of measurements and the short term coverage allowed the use of 
DFT, while the number of data points was insufficient to calculate the pulsation 
period properly with PDM. Since the time interval of the \emph{Gaia} 
photometric data was covered by ASAS-SN and KWS observations, the period 
was calculated with the help of $O-C$ method as well. The photometric data 
obtained by \emph{Hipparcos} had to be treated differently, since the 
measurements covered only short time, but the amount of data points was 
larger than in the case of \emph{Gaia} measurements. This time interval 
was not covered by any other surveys, thus the pulsation period was only 
calculated with DFT and PDM for this survey. In the case of the data 
obtained at Piszk\'estet\H{o} Observatory, the measurements were scattered 
in time and covered several years. For this reason, the pulsation period 
was only calculated with $O-C$ method. The final period value was the 
average of the pulsation periods obtained with the different methods for 
every survey (Table \ref{tab:period_tab}). The pulsation period of AU~Peg 
is presented as a function of time in Fig.~\ref{fig:periods}.

According to the \emph{Hipparcos} and Piszk\'estet\H{o} measurements, 
the pulsation period of AU~Pegasi was slightly increasing between 
HJD\,2\,448\,500 and 2\,452\,000 at a rate of 8.348$\cdot 10^{-5}$ day/year. 
Between HJD\,2\,452\,900 and 2\,454\,850 the rate of the period change 
increased according to the ASAS measurements, to the value of 
3.746$\cdot 10^{-4}$ day/year, which is approximately half of the rate 
the pulsation period had been changing with between HJD\,2\,442\,500 and 
2\,446\,000. After this rapid change, the period seemed to keep its value, 
and it remained constant until the latest observations. This behaviour 
has not been observed before and it is still an open question, how the 
companion of the Cepheid affects the pulsation, and if there is a direct 
connection between the evolution of the period and binarity of the star.

\begin{table}[!h]
\begin{center}
\scalebox{0.88}{
\begin{tabular}{c  c c c c c}
\hline
Survey & $T$ & P$_{\textrm{DFT}}$ & P$_{\textrm{PDM}}$ & P$_{O-C}$ & P$_{\textrm{final}}$ \\
\hline
ASAS & 52900 & 2.4103 & 2.4105 & 2.4111 & 2.4106 \\
 & & \hspace*{-4pt}$\pm 0.0002$ & \hspace*{-4pt}$\pm 0.0001$ & \hspace*{-4pt}$\pm 0.0001$ & \hspace*{-4pt}$\pm 0.0003$ \\
 & 53600 & 2.4119 & 2.4119 & 2.4111 & 2.4117  \\
 & & \hspace*{-4pt}$\pm 0.0001$ & \hspace*{-4pt}$\pm 0.0001$ & \hspace*{-4pt}$\pm 0.0001$ & \hspace*{-4pt}$\pm 0.0003$ \\
 & 54300 & 2.4133 & 2.4130 & 2.4128 & 2.4130 \\ 
 & & \hspace*{-4pt}$\pm 0.0001$ & \hspace*{-4pt}$\pm 0.0002$ & \hspace*{-4pt}$\pm 0.0001$& \hspace*{-4pt}$\pm 0.0002$ \\
 & 54800 & 2.4124 & 2.4126 & 2.4128 & 2.4126 \\
  & & \hspace*{-4pt}$\pm 0.0009$ & \hspace*{-4pt}$\pm 0.0003$ & \hspace*{-4pt}$\pm 0.0001$& \hspace*{-4pt}$\pm 0.0004$ \\

ASAS-SN & 56600 & 2.4128 & 2.4130 & 2.4128 & 2.4129 \\
 & & \hspace*{-4pt}$\pm 0.0007$ & \hspace*{-4pt}$\pm 0.0003$ & \hspace*{-4pt}$\pm 0.0001$& \hspace*{-4pt}$\pm 0.0002$ \\
 & 56900 & 2.4128 & 2.4130 & 2.4128 & 2.4129 \\
  & & \hspace*{-4pt}$\pm 0.0008$ & \hspace*{-4pt}$\pm 0.0002$ & \hspace*{-4pt}$\pm 0.0001$& \hspace*{-4pt}$\pm 0.0002$ \\
 & 57200 & 2.4129 & 2.4132 & 2.4128 & 2.4130 \\
  & & \hspace*{-4pt}$\pm 0.0004$ & \hspace*{-4pt}$\pm 0.0001$ & \hspace*{-4pt}$\pm 0.0001$& \hspace*{-4pt}$\pm 0.0002$ \\
 & 57500 & 2.4127 & 2.4134 & 2.4128 & 2.4130 \\
  & & \hspace*{-4pt}$\pm 0.0002$ & \hspace*{-4pt}$\pm 0.0002$ & \hspace*{-4pt}$\pm 0.0001$& \hspace*{-4pt}$\pm 0.0002$ \\
 & 58000 & 2.4126 & 2.4130 & 2.4128 & 2.4128 \\
  & & \hspace*{-4pt}$\pm 0.0009$ & \hspace*{-4pt}$\pm 0.0002$ & \hspace*{-4pt}$\pm 0.0001$& \hspace*{-4pt}$\pm 0.0002$ \\
 & 58300 & 2.4127 & 2.4128 & 2.4128 & 2.4128 \\
  & & \hspace*{-4pt}$\pm 0.0004$ & \hspace*{-4pt}$\pm 0.0001$ & \hspace*{-4pt}$\pm 0.0001$& \hspace*{-4pt}$\pm 0.0001$ \\

\emph{Gaia} & 57200 & 2.4128 & $-$ & 2.4128 & 2.4128 \\
 & & \hspace*{-4pt}$\pm 0.0003$ &  & \hspace*{-4pt}$\pm 0.0001$ & \hspace*{-4pt}$\pm 0.0001$ \\

\emph{Hipparcos} & 48500 & 2.4103 & 2.4104 & $-$ & 2.4104 \\
 & & \hspace*{-4pt}$\pm 0.0002$ & \hspace*{-4pt}$\pm 0.0001$ & & \hspace*{-4pt}$\pm 0.0001$ \\

KWS & 55900 & 2.4131 & 2.4130 & 2.4128 & 2.4130 \\
 & & \hspace*{-4pt}$\pm 0.0002$ & \hspace*{-4pt}$\pm 0.0001$ & \hspace*{-4pt}$\pm 0.0001$ & \hspace*{-4pt}$\pm 0.0001$ \\
 & 56700 & 2.4127 & 2.4130 & 2.4128 & 2.4128 \\
  & & \hspace*{-4pt}$\pm 0.0003$ & \hspace*{-4pt}$\pm 0.0001$ & \hspace*{-4pt}$\pm 0.0001$ & \hspace*{-4pt}$\pm 0.0002$ \\
 & 57500 & 2.4127 & 2.4129 & 2.4128 & 2.4128 \\
  & & \hspace*{-4pt}$\pm 0.0003$ & \hspace*{-4pt}$\pm 0.0001$ & \hspace*{-4pt}$\pm 0.0001$ & \hspace*{-4pt}$\pm 0.0001$ \\
 & 58100 & 2.4129 & 2.4129 & 2.4128 & 2.4129 \\
  & & \hspace*{-4pt}$\pm 0.0002$ & \hspace*{-4pt}$\pm 0.0002$ & \hspace*{-4pt}$\pm 0.0001$ & \hspace*{-4pt}$\pm 0.0001$ \\

Present  & 49600 & 2.4110 & $-$ & 2.4109 & 2.4109 \\
 paper& & \hspace*{-4pt}$\pm 0.0021$ &  & \hspace*{-4pt}$\pm 0.0001$ & \hspace*{-4pt}$\pm 0.0006$ \\
 & 50800 & 2.4113 & $-$ & 2.4111 & 2.4111 \\
  & & \hspace*{-4pt}$\pm 0.0033$ &   & \hspace*{-4pt}$\pm 0.0001$ & \hspace*{-4pt}$\pm 0.0010$ \\
 & 52000 & 2.4112 & $-$ & 2.4111 & 2.4111 \\
  & & \hspace*{-4pt}$\pm 0.0002$ &   & \hspace*{-4pt}$\pm 0.0001$ & \hspace*{-4pt}$\pm 0.0002$ \\
\hline
\end{tabular}
}
\caption{The pulsational period values calculated from the seasonal 
data of different surveys. $T$: HJD$-$2\,400\,000; $P$: the obtained 
period in days.}
\label{tab:period_tab}
\end{center}
\end{table}

\section{Revised spectroscopic orbit of AU~Peg}

The orbital elements of the binary system involving AU~Peg have been
determined by Harris et~al. (1984) based on their own RV data. When
revising the elements of the spectroscopic orbit, the available RV 
data (i.e. those published by \cite{Harris1984}, \cite{Barnes1988},
\cite{Gorynya1998} and \cite{Vinko1998}) were split into subsets 
covering no more than two years. These data sets then have been 
corrected for variations due to the pulsation by fitting and subtracting 
the sinusoidal changes corresponding to the pulsation period and its first
harmonic. 
The amplitude ratio of the two fitted components is 10:1. 
The second harmonic can be neglected, since its amplitude is not large enough 
to make the signal distinguishable from the noise in the Fourier spectrum. 
Since the pulsation period of the Cepheid was changing 
significantly during and between the different RV measurements, we used 
the diagram shown in Fig.~\ref{fig:periods} to obtain the correct value 
for the pulsation period. While subtracting the contribution of the first
harmonic from the RV data, we assumed that the RV amplitude ratio of the
fundamental and first harmonic variation is the same as in the case of 
the light curve. To obtain the orbital parameters we fitted 
$$v_i = V - K(\cos{(f_i + \overline{\omega})} + e \cos{\overline{\omega}}) $$
to the pulsation corrected dataset, where $v_i$ denotes the $i$th RV 
entry corresponding to $f_i$ true anomaly at time entry $t_i$. 
In the formula above, $V$ is the systemic velocity of the system, $K$ 
is the semi-amplitude of the variation, $e$ is the eccentricity 
of the orbit, and $\overline{\omega}$ is the argument of the periapsis 
\citep{Fulton2018}. To calculate the mean anomalies at various time entries 
we also had to fit the periastron passage factor $\chi$, which is the fraction 
of orbit prior to the start of data-taking that periastron occurred.

\begin{figure}[!h]
\centering
\hspace*{-13pt}
\includegraphics[scale=0.36]{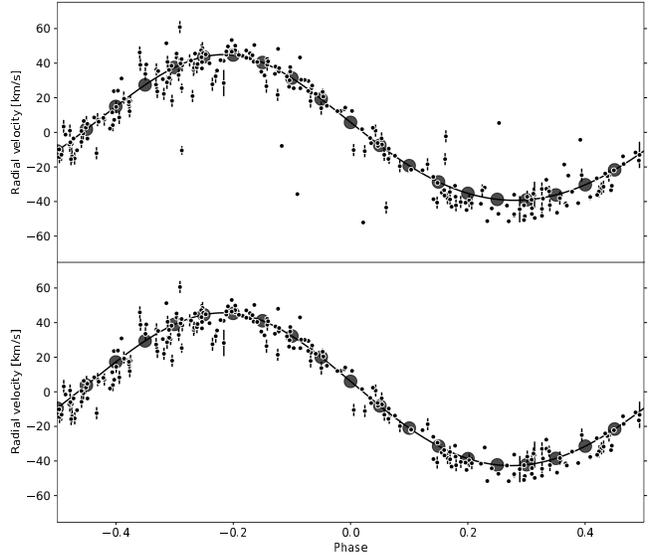}
\caption{The orbital RV curve (after correcting for the pulsation) and
the obtained fit. Top panel: the obtained fit with the discrepant 
datapoints included. Bottom panel: the fit obtained after removing these points.}
\label{fig:rad_fit}
\end{figure}

We used a Bayesian approach to fit the RV data and extracted the orbital 
parameters along with their uncertainties using Markov Chain Monte Carlo 
(MCMC) simulations. To implement this method we have used the 
\texttt{radvel} python package introduced and described in 
\cite{Fulton2018}. The prior distributions were chosen to be uniform 
centered on the parameters obtained in \cite{Harris1984}, except for 
the eccentricity and the longitude of the periastron, for which every 
possible value was considered. The obtained fit along with the orbital 
RV phase curve is presented in Fig. \ref{fig:rad_fit}.

The computed elements are 
\begin{align*}
P &= 53.3344 \pm 0.0003 \textrm{ d} \\
V &= -1.6368 \pm 0.0004 \textrm{ km/s} \\
K &= 44.091 \pm 0.592 \textrm{ km/s} \\
e &= 0.0425 \pm 0.0027 \\
\chi &= 0.3264 \pm 0.0005 \\
\overline{\omega} &= 0.3404 \pm 0.0040 \textrm{ rad}.
\end{align*}
We have compared these orbital elements to those obtained in 
\cite{Harris1984}. According to our study, the orbit is fairly 
different from the previously assumed one: the orbital period 
appears to be longer than obtained before ($53.319 \pm 0.015$ 
days) and the calculated amplitude of variation is larger by 
2.2 km/s than the previous one, as well. The eccentricity of the 
orbit appears to be smaller, thus according to our calculations, 
the orbit itself is more circular, than it was originally believed 
($e = 0.12 \pm 0.04$, \cite{Harris1984}). It has been mentioned in 
\cite{Harris1984}, that by omitting a discrepant point from their 
dataset, they obtained an orbit with smaller eccentricity, which 
is more similar to the solution we obtained. Since for the solution 
above we discarded all discrepant points (through creating the phase curve in every 250 day long 
time interval supposing a sinusoidal change, then removing the outliers 
with the help of the previously shown $2\sigma$ clipping) we tested whether the 
orbit obtained from the original data, including the previously 
deleted points, would be more similar to the one in \cite{Harris1984}. 
In this case the period remained the same, but the amplitude 
decreased by 2 km/s, thus its value became very similar to that  
obtained in \cite{Harris1984}. The eccentricity of the orbit turned out 
to be larger than in the first case ($e = 0.068 \pm 0.004$)
due to these discrepant datapoints. Although this value is within the 
error limits given in \cite{Harris1984}, it still corresponds to a 
more circular orbit and we believe, that omitting the discrepant 
datapoints is a reasonable choice, considering their high scatter 
from the fitted phase curve (Fig.~\ref{fig:rad_fit}.).

\section{Summary}

It has been presented that, in contrast to the previous behaviour of 
AU~Pegasi, the rate of pulsation period change has decreased significantly 
and the period has come to a constant value, according to the latest 
observations. The last data point from the analysis of \cite{Vinko1993} 
suggests that there might have been another time interval (between 
JD\,2\,445\,000 and 2\,447\,500), when the pulsation period set in a 
constant value over time, followed by a rapid period change. If this 
behaviour turns out to be periodic, it might be an indicator for the 
interaction between the Cepheid and its companion.

According to our analysis, a wave in the $O-C$ diagram has been 
found, which could not have been connected to any known physical 
process in the environment of the Cepheid. The amplitude and period 
of this variation might correspond well to light-time effect, 
but the lack of this periodicity in the RV data rules this 
possibility out. This effect might be the result of the tidal 
interaction between the Cepheid and its companion, but to 
support this hypothesis, further observations and analysis 
would be required.

We have also revised the spectroscopic orbit of AU~Peg. By subtracting 
the contributions of the pulsation from the RV data taking into account
the strong changes observed in the pulsation period (see Fig.~7)
and by using Bayesian framework for the fitting, we could reliably 
determine the orbital elements of AU~Peg. According to our analysis, 
the orbit of AU~Peg is more circular than it was previously determined, 
regardless how one handles the discrepant datapoints (the eccentricity 
obtained in the case of the whole dataset is $e=0.068 \pm 0.004$, while 
after removing the mentioned datapoints the fitting process resulted in 
$e = 0.0425 \pm 0.0027$). The resulting amplitude of the RV variation 
and the orbital period values were larger than the ones obtained in 
\cite{Harris1984}, which, together with the smaller eccentricity, 
indicate higher mass function for the companion star.

The peculiar behaviour of the pulsation period of AU~Pegasi necessitates
frequent photometric observations of this interesting binary system with
a Type~II Cepheid component.

\acknowledgments
This project has been supported by the GINOP-2.3.2-15-2016-00003 grant 
of the Hungarian National Research, Development and Innovation Office 
(NKFIH). This work has been partly supported by the Lend\"ulet Program 
of the Hungarian Academy of Sciences, project No. LP2018-7/2018 and the 
Hungarian NKFIH projects K-115\,709 and K-129\,249.

\newpage

\bibliographystyle{spr-mp-nameyear-cnd}
\bibliography{Bibliography}

\end{document}